\documentclass{pasj01}
\draft 
\Received{$\langle$April 03, 2018$\rangle$}
\Accepted{$\langle$July 21, 2018$\rangle$}
\Published{$\langle$publication date$\rangle$}

\usepackage{color}
\usepackage{ifthen}

\newcommand{\fscat}{\ensuremath{F_{\rm{scat}}}}

\newcommand{\nh}{\ensuremath{N_\mathrm{H}}}
\newcommand{\ledd}{\ensuremath{L_\mathrm{edd}}}
\newcommand{\lpeak}{\ensuremath{L_\mathrm{peak}}}

\newcommand{\erg}{\mathrm{erg}\ \mathrm{cm}^{-2}\ \mathrm{s}^{-1}}

\newcommand{\Tin}{T_\mathrm{in}}

\newcommand{\Rin}{R_\mathrm{in}}

\newcommand{\msolar}{M_{\odot}}

\usepackage{color}

\usepackage{xcolor}
\usepackage{ulem}
\definecolor{ins}{rgb}{0.5, 0, 0}
\definecolor{del}{rgb}{0, 0, 0.5}

\newcommand{{\maxi}}{\rm MAXI}
\newcommand{{\swift}}{\it Swift}

\newcommand{{\np}}{114} 

\begin{document}

\SetRunningHead{Nakahira et al.}{Study of mass accretion in MAXI J1535$-$571}

\title{Discovery and state transitions of the new Galactic black hole candidate MAXI J1535$-$571}

\author{
	Satoshi \textsc{Nakahira}\altaffilmark{1}, 
	Megumi \textsc{Shidatsu}\altaffilmark{1,2}, 
	Kazuo \textsc{Makishima}\altaffilmark{1}, 
  	Yoshihiro \textsc{Ueda}\altaffilmark{3},
 	Kazutaka \textsc{Yamaoka}\altaffilmark{4},
	Tatehiro \textsc{Mihara}\altaffilmark{1}, 
	Hitoshi \textsc{Negoro}\altaffilmark{5}, 
	Tomofumi \textsc{Kawase}\altaffilmark{5}, 
	Nobuyuki \textsc{Kawai}\altaffilmark{6}, 
 and 
	Kotaro \textsc{Morita}\altaffilmark{6}
	}

\altaffiltext{1}{MAXI team, RIKEN, 2-1 Hirosawa, Wako, Saitama 351-0198, JAPAN}
\altaffiltext{2}{Department of Physics, Ehime University, 
2-5, Bunkyocho, Matsuyama, Ehime 790-8577, Japan}
\altaffiltext{3}{Department of Astronomy, Kyoto University, Kitashirakawa-Oiwake-cho, Sakyo-ku, Kyoto 606-8502, Japan}
\altaffiltext{4}{Institute for Space-Earth Environmental Research (ISEE), Nagoya University, Furo-cho, Chikusa-ku, Nagoya, Aichi 464-8601, Japan}
\altaffiltext{5}{Department of Physics, Nihon University, 1-8 Kanda-Surugadai, Chiyoda-ku, Tokyo 101-8308, Japan}
\altaffiltext{6}{Department of Physics, Tokyo Institute of Technology, 2-12-1 Ookayama, Meguro-ku, Tokyo 152-8551, Japan}

%


\maketitle

\begin{abstract}

We report on the detection and subsequent X-ray monitoring of the
new Galactic black hole candidate MAXI J1535$-$571 with the MAXI/GSC. After
the discovery on 2017 September 2 made independently with $\maxi$ and 
the $\swift$/BAT, the source brightened gradually, and in a few weeks, 
reached the peak intensity of $\sim$ 5 Crab, or  
$\sim 1.6 \times 10^{-7}~\erg$ in terms of the 2--20 keV flux. 
On the initial outburst rise, the 2-20 keV $\maxi$/GSC spectrum was described by 
a power-law model with a photon index of $\lesssim 2$, while 
after a hard-to-soft transition which occurred on September 18, 
the spectrum required a disk blackbody component in addition.
At around the flux peak, the 2-8 keV and 15-50 keV light curves 
showed quasi-periodic and anti-correlated fluctuations with an amplitude of 10--20 \%,
on a time scale of $\sim$1-day.
Based on these X-ray properties obtained with the $\maxi$/GSC, 
with additional information from the $\swift$/BAT, 
we discuss the evolution of the spectral state of this source, 
and give constraints on its system parameters.

\if 1
Discovery and state transitions of the new Galactic black hole candidate MAXI J1535-571

Satoshi NAKAHIRA; Megumi SHIDATSU1; Kazuo MAKISHIMA;  Yoshihiro UEDA; Kazutaka YAMAOKA; Tatehiro MIHARA; Hitoshi NEGORO; Tomofumi KAWASE; Nobuyuki KAWAI; and Kotaro MORITA

\fi

\end {abstract}

\section{introduction}\label{sec:introduction}

Transient black hole binaries (BHBs) sometimes exhibit dramatic outbursts, changing its X-ray luminosity over several orders of magnitude. 
Their outburst light curves typically consist of a rapid rise on timescales of a few days to a few weeks, and a subsequent slow decay on timescales of weeks to a few months. 
In their outbursts, they show distinct ``states'' with different spectral properties (e.g., \cite{2006csxs.book..157M, Done:2007hu}). The canonical two states are the so-called hard state and the soft state, where the spectrum below 10 keV is characterized by a hard power-law component with a photon index of $<2$, and by a multi-color disk blackbody component \citep{Mitsuda:1984ul}, respectively. 
BHBs are known to exhibit a hysteresis in their flux versus spectral evolution; they make state transitions from the hard state to the soft state and the opposite way at different luminosities, and consequently track a q-shaped curve on an X-ray hardness-intensity diagram (HID; \cite{2005ApSS.300..107H}).  
Because the spectral properties are considered to reflect the structure and dynamics of accretion flows in the vicinity of the central black holes, probing the complex spectral evolution during outbursts gives us clues to the physics of black hole accretion over a wide range of mass accretion rates.

MAXI J1535$-$571 was detected on 2017 September 2 with the GSC onboard MAXI \citep{2009PASJ...61..999M}, 
which is operating on the International Space Station (ISS). 
The $\maxi$/GSC nova alert system \citep{Negoro:2016jb} triggered on the source at UT 14:41 
with a low significance. 
As reported by \citet{2017ATel10699....1N}, the detection was confirmed with subsequent 
scans, which recorded an X-ray intensity of $34\pm6$ mCrab in 4--10 keV. 
Almost simultaneously and independently, the Swift/BAT also discovered the source 
\citep{2017GCN.21788....1M, 2017ATel10700....1K}, and precisely determined its position at ($\alpha$, $\delta$) $=$ (15$^\mathrm{h}$ 35$^\mathrm{m}$ 19.73$^\mathrm{s}$, -57$^\circ$ 13$'$ 48$''$.1) with the XRT and UVOT onboard. After the discoveries and a gradual increase of the flux, the source spectrum softened on about September 18 \citep{2017ATel10729....1N,2017ATel10731....1K, 2017ATel10733....1P} and a disk blackbody component became significant \citep{2017ATel10761....1S}, suggesting a state transition from the hard state into the soft state. 

Follow-up observations of the object were extensively 
performed in various wavelengths \citep{2017ATel10702....1S, 2017ATel10899....1R, 2017ATel10716....1D, 2017ATel10729....1N, 2017ATel10761....1S, 2017ATel10731....1K, 2017ATel10734....1M, 2017ATel10711....1R, 2017ATel10768....1G, 2017ATel10745....1T, 2017ATel10816....1B}. 
The counterpart of MAXI J1535$-$571 was found in 
the optical and infrared bands soon after the onset of the outburst, 
with a relatively low optical flux ($\sim$21 mag in the $i'$ band; 
\cite{2017ATel10702....1S, 2017ATel10716....1D}). Radio emission from 
the object was also detected \citep{2017ATel10711....1R,2017ATel10745....1T}. 
The source was interpreted as a new BH candidate in our galaxy, 
considering the X-ray spectral shapes and rapid X-ray variability 
\citep{2017ATel10708....1N}, as well as the radio versus X-ray flux ratio 
\citep{2017ATel10899....1R}. The optical/near-infrared variations 
suggest that the companion is a low-mass star 
\citep{2017ATel10702....1S, 2017ATel10716....1D}.
The NuSTAR spectrum 
taken on the 5th day from the outburst onset suggests a low electron temperature, $\sim$20 keV, and a high spin of $a >$0.84 assuming the source to be an accreting BH in the hard state \citep{2018ApJ...852L..34X}. 

As a working hypothesis, we here assume that the object is a new BHB.
Based manly on the ${\maxi}$/GSC data, the present paper describes 
the X-ray behavior of MAXI J1535$-$571 over the 220 days from the outburst onset.

\section{Data analysis and results}
\subsection{Data reduction}
We studied the X-ray properties of MAXI J1535$-$571 using 
light curves and spectra obtained with the $\maxi$/GSC.
The 15--50 keV light curve of the Swift/BAT was also employed 
to investigate the long-term trend in the hard 
X-ray band.

We analyzed the $\maxi$ data with the processed version 1.3.6.6, 
which is a beta-test version released from DARTS at 
ISAS/JAXA\footnote{http://darts.isas.jaxa.jp/astro/maxi/}.
Light curves and spectral data were produced from the event data 
via $\maxi$ specific tools implemented in ``MAXI/GSC on-demand web 
interface \footnote{http://maxi.riken.jp/mxondem}\citep{Nakahira:2013we}''.  
The on-source data were extracted from a circular region with 
a radius of 2$^\circ$.1, centered on the source position. The background data 
of the same region were produced with a background event generator, which 
adopts the same method as used in the second MAXI/GSC extragalactic source catalog 
\citep{Hiroi:2013gt}. To suppress statistical fluctuations, we generated 100 
times more background counts than in the real data.

For the Swift/BAT data, we used the archived 15--50 keV light curve  
with a time resolution of $\sim 92$ minutes (the spacecraft orbital period), 
available on the ``BAT Transient Monitor'' website (\cite{2013ApJS..209...14K}) \footnote{http://swift.gsfc.nasa.gov/docs/swift/results/transients}.

\subsection{Light curves and hardness ratios}\label{sec:lightcurve}

\begin{figure*}
  \begin{center}
    \FigureFile(165mm,165mm){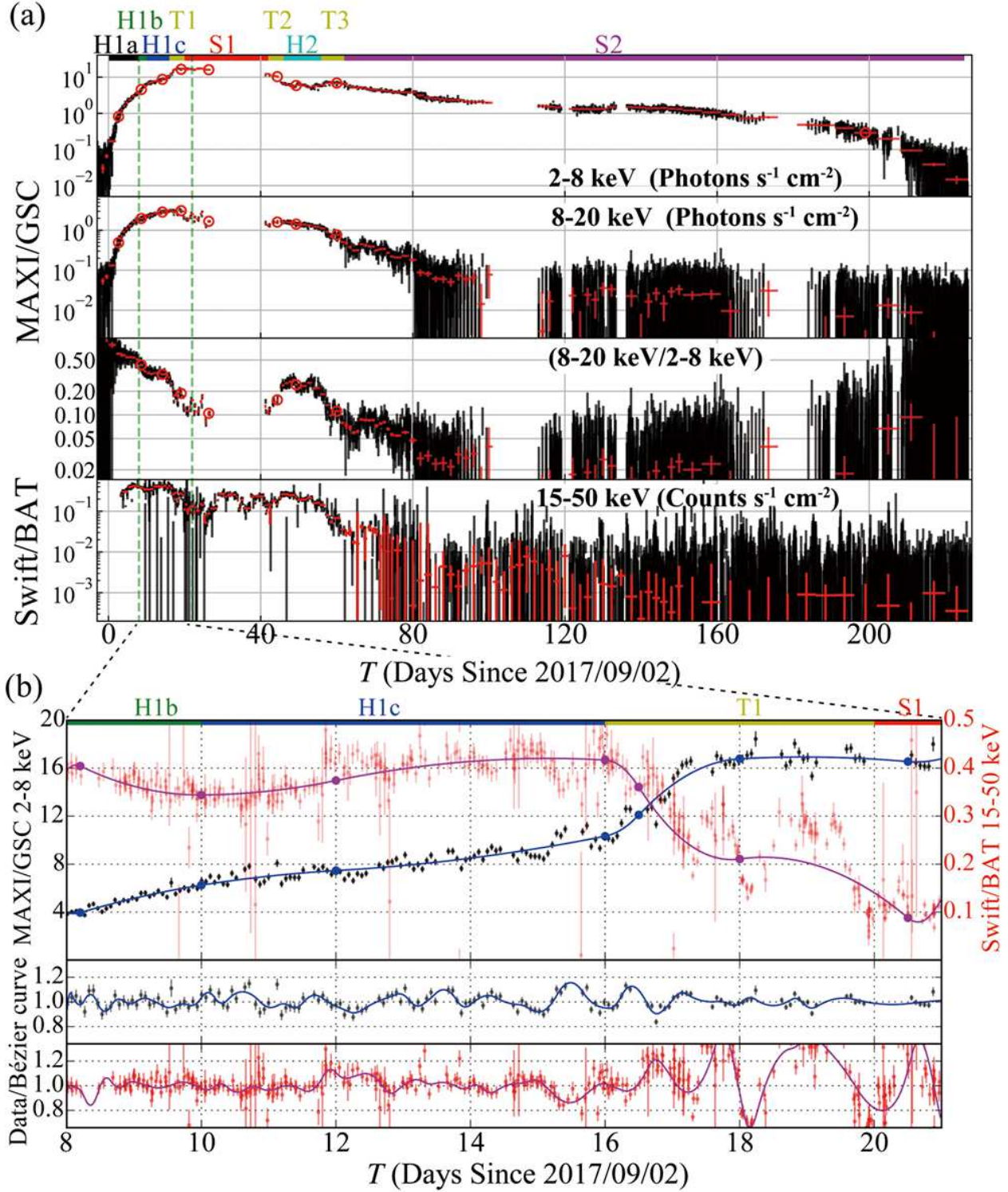}
  \end{center}
  \caption{
  (a) The {\maxi}/GSC 2--8 keV and 8--20 keV light curves, 
  hardness ratio between 8--20 keV and 2--8 keV band, and the 
  15--50 keV {\swift}/BAT light curve, from top to bottom.
  Black points indicate data with the finest time resolution  ($\sim$92 minutes), 
  and red points are binned data.
  (b) Top: the zoomed 2--8 keV (black) and 15--50 keV (red) light curves 
   over the period indicated by a pair of green lines in panel(a), 
   which includes the first hard-to-soft state transition. 
  The second order B$\acute{e}$zier curves (solid 
  lines) and its knots (circles) are superposed on the data. 
  Middle and bottom: the ratios of the data to the B$\acute{e}$zier curve.

  Errors in all panels represent 1$\sigma$ confidence intervals.
   }
\label{fig:lightcurve}
\end{figure*}

Figure~\ref{fig:lightcurve}a shows background-subtracted MAXI/GSC light curves of MAXI J1535$-$571 in 2--8 keV and 8--20 keV, and their hardness ratio (HR). The time origin ($T$=0 day) was chosen at the onset date of the outburst ( = 2017 September 2 ).
Each data point represents one scan transit of the source, which takes place 
every 92 minutes and lasts for 40-100 s.
$\maxi$ was not able to observe 
the direction of MAXI J1535$-$571 for $T=$25--42, $T=$101--112, 120--121 or 173-182.
A Swift/BAT light curve in 15--50 keV is also presented to compare intensity variations 
at different energies. 
Because the data with the original sampling are subject to large statistical errors, 
we took averages typically over 2--20 adjacent data points.
These binned data, $\np$ points in total, are show in red in figure~\ref{fig:lightcurve}a.

\begin{figure}
  \begin{center}
    \FigureFile(80mm,80mm){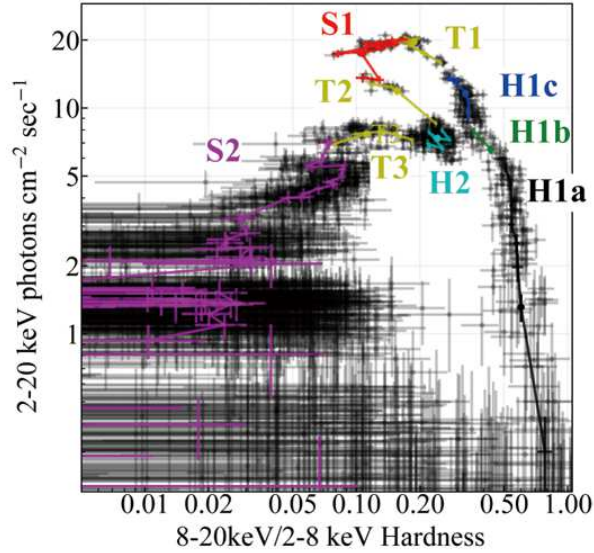}
  \end{center}
  \caption{
  A scatter plot, namely an HID, between the hardness ratio vs. 2--20 keV intensity of the MAXI/GSC data with the 92-min time resolution. 
  Color points connected with lines are binned data produced in the same manner as in figure\ref{fig:lightcurve}a.}
\label{fig:HID}
\end{figure}

The MAXI/GSC light curves and their HRs are converted to an HID in 
figure~\ref{fig:HID}. There, colored data points represent the binned data in  figure~\ref{fig:lightcurve}.
Considering the behavior of the light curves and HID with MAXI, 
the entire outburst period can be classified into the following nine phases.
Among them, 
H1 and H2 represent relatively hard phase typical of the hard state, 
S1 and S2 relatively soft phases to be interpreted as the soft state, 
and T1-T3 transient periods.

\begin{description}
\item[H1a ($T$=0--8):] the 2--8 and 8--20 keV intensities steadily increased, keeping an approximately constant HR $\sim$0.6. 
\item[H1b ($T$=8--10):] the 2--8 keV flux increased more rapidly than in H1a, and accordingly the HR rapidly declined to $\sim$0.35. 
\item[H1c ($T$=10--16):] the fluxes in both bands again increased steadily, with a slight decrease of the HR. 
\item[T1 ($T$=16--20):] the source declined rapidly in 2-8 keV and brightened in 8-20 keV simultaneously; this is regarded as the first hard-to-soft state transition.
\item[S1 ($T$=20--42):] the 2--8 keV flux was kept at the maximum level, while the 8--20 keV flux gradually decreased with small fluctuations (see below).
The HR reached 0.1--0.2, the lowest value of the first portion in the HID. 
\item[T2 ($T$=42--46):] the HR rapidly increased, and the source returned to the hard state.
\item[H2 ($T$=46--56):] the HR remained relatively constant, and the source made the second hard-to-soft state transition.
\item[T3 ($T$=56--62):] the HR rapidly decreased.
\item[S2 ($T$=62--):] the 2--8 and 8--20 keV fluxes both decreased steadily, while the HR marked the minimum value in the outburst, $\sim 0.01$. Then, after $T$ = 200, the source declined more rapidly, keeping the soft spectrum.

\end{description}
The behavior of the 15--50 keV flux was different from that in 8--20 keV; 
it decreased in H1b and then rose again in H1c. After the rapid decrease 
at the beginning of T1, 
it jumped among a few rather discrete intensity levels, 
and then gradually decreased during H2, T3 and S2 to undetectable levels.

Around $T$=8--20, the $\maxi$ and $\swift$ data were found to scatter much more than the 
statistical errors on a timescale of $\sim$one day. Figure~\ref{fig:lightcurve}b 
shows an expanded light curve for $T$=8--22, where superposed is a 
smoothed light curve produced using the second order B$\acute{e}$zier curve. 
Anti-correlated oscillations can be seen between the 2--8 keV and 15--50 keV 
intensities with an amplitude of $\sim$10--20\% and a period of $\sim$ one day.

\subsection{Energy spectra}\label{sec:spectrum}

We extracted time-averaged MAXI/GSC spectra of MAXI J1535$-$571 from 
the $\np$ binned time points (defined in Section~\ref{sec:lightcurve}).
The derived spectra were then fitted with the standard X-ray emission 
model for BH X-ray binaries: a disk blackbody emission 
and its Comptonization, both absorbed by cold interstellar plus circum-source medium. 
We adopted the multi-color disk model {\tt diskbb} 
\citep{Mitsuda:1984ul}, and convolved it with {\tt simpl} 
\citep{Steiner:2009gh} in which a fraction of the input 
seed photons are redistributed by Comptonization into a power-law form. 
The absorption was expressed by the {\tt TBabs}  model,
referring to the solar-abundance table given by \citet{Wilms:2000en}. 
Because {\tt simpl} is a convolution model, the energy band used 
in the spectral fitting was extended down to 0.01 keV and up to 100 keV. 
The spectral analysis was carried out with XSPEC 
version 12.9.1, and the errors represent 90\% confidence limits.

In the fitting, the absorption columns density was fixed at $\nh$=2.6$\times$
10$^{22}$ cm$^{2}$, a value which was favored by essentially all the spectra.
We fixed $\Gamma$ at 2.50, a typical value during the soft state 
of BHBs \citep{2006csxs.book..157M}, 
because the GSC energy band was dominated by the direct disk emission and hence 
$\Gamma$ of the {\tt simpl} model was poorly constrained.
We confirmed that adopting $\Gamma$ from 2.2 to 2.7, instead of 
2.50, does not affect our results.
When the HR (8--20 keV vs 2--8 keV) is higher than 0.20, 
a single {\tt powerlaw} model with free $\Gamma$ was used instead 
of the {\tt simpl*diskbb} model, because in there spectra the {\tt diskbb} 
component cannot be constrained by the GSC data. 

The inner disk radius $R_\mathrm{in}$ was estimated from the 
normalization of the {\tt diskbb} component, by 
applying a correction factor of $\xi\kappa^{2}=$1.18, which is obtained by 
combining an adjustment to express stress-free boundary condition 
($\xi$ = 0.41: see e.g., \cite{1998PASJ...50..667K})
with the color hardening factor ($\kappa$ = 1.7; 
see e.g., \cite{Shimura:1995te}).

\begin{figure*}
  \begin{center}
    \FigureFile(55mm, 55mm){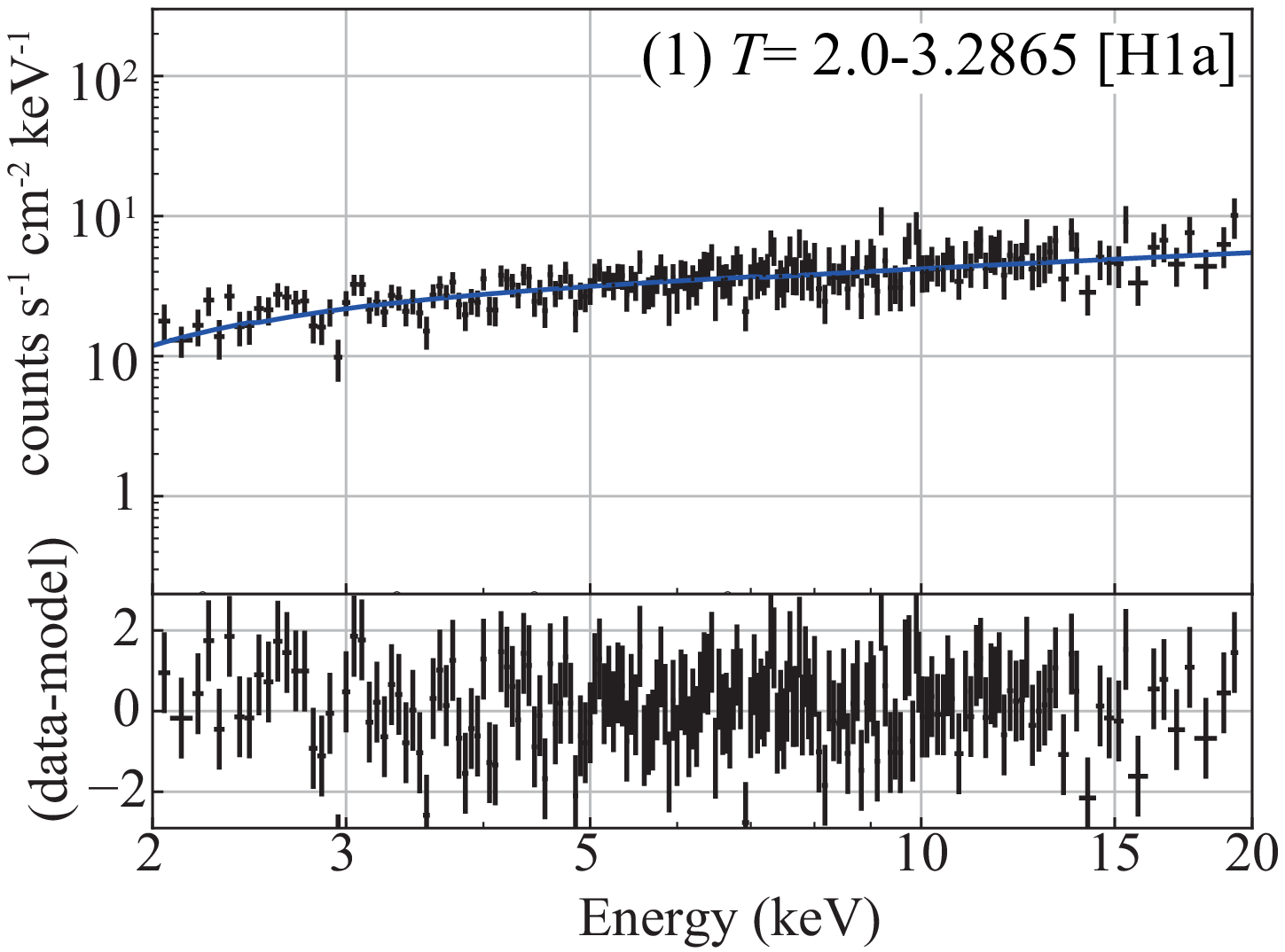}
    \FigureFile(55mm, 55mm){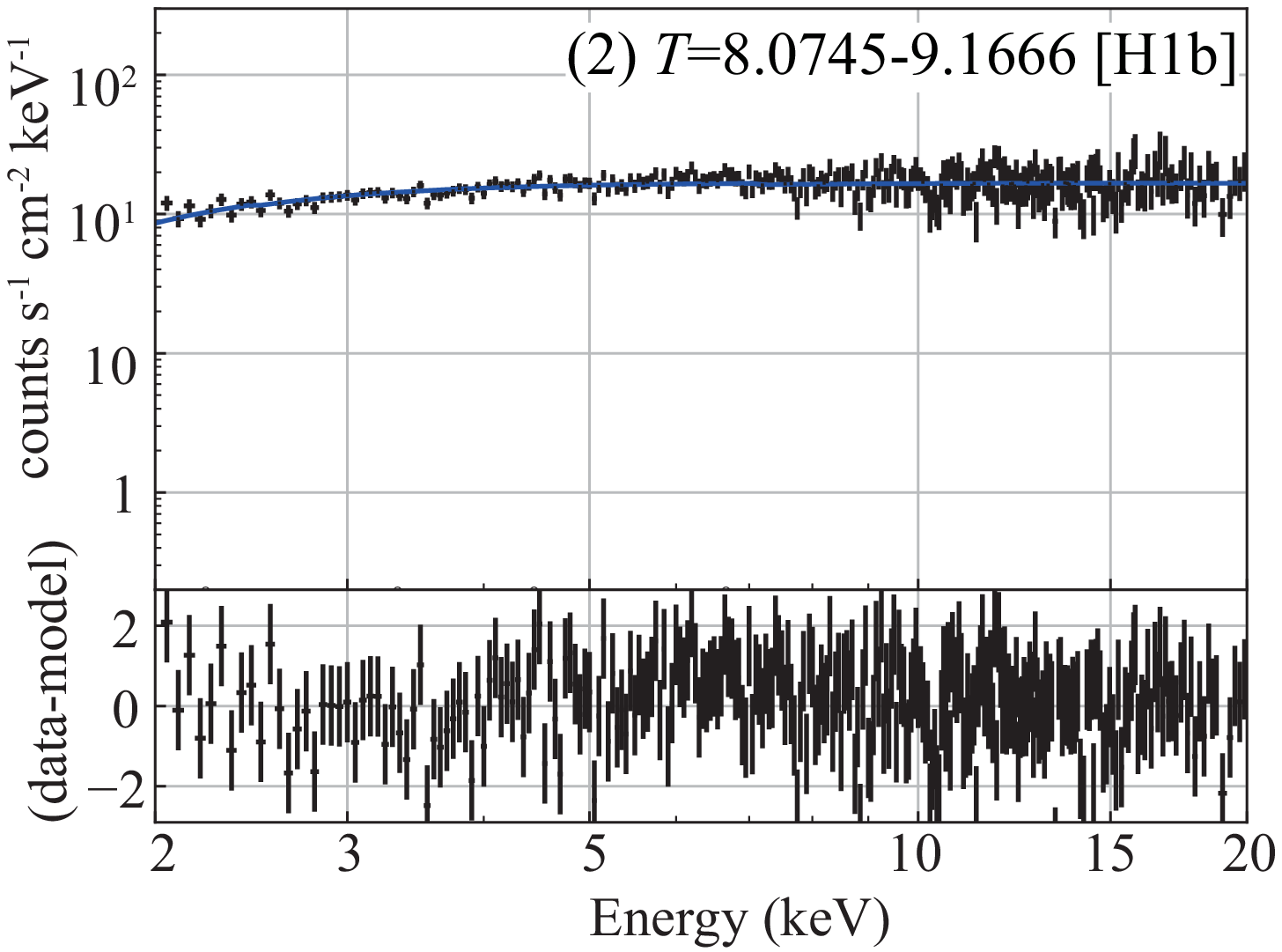}
    \FigureFile(55mm, 55mm){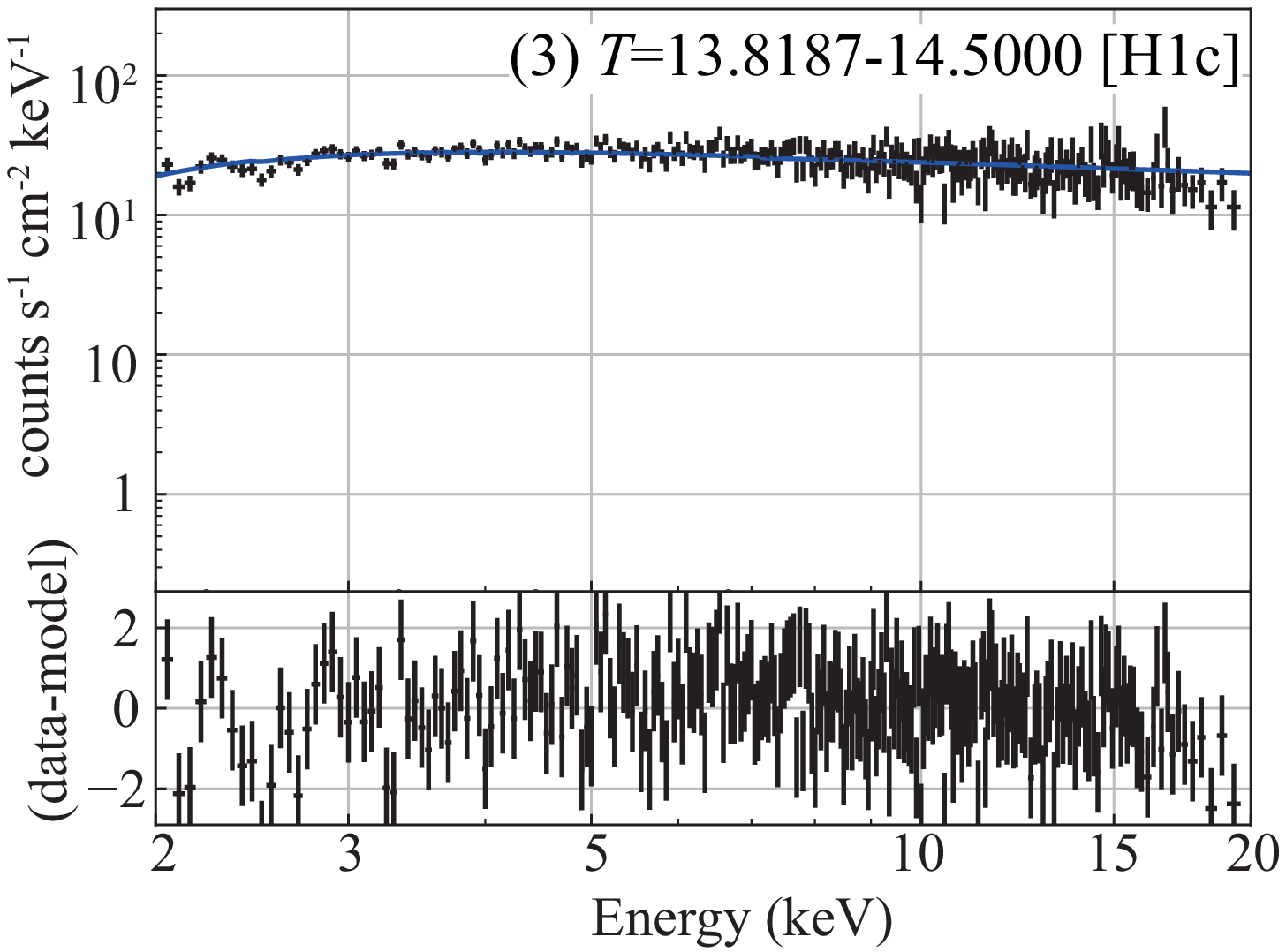}
    \FigureFile(55mm, 55mm){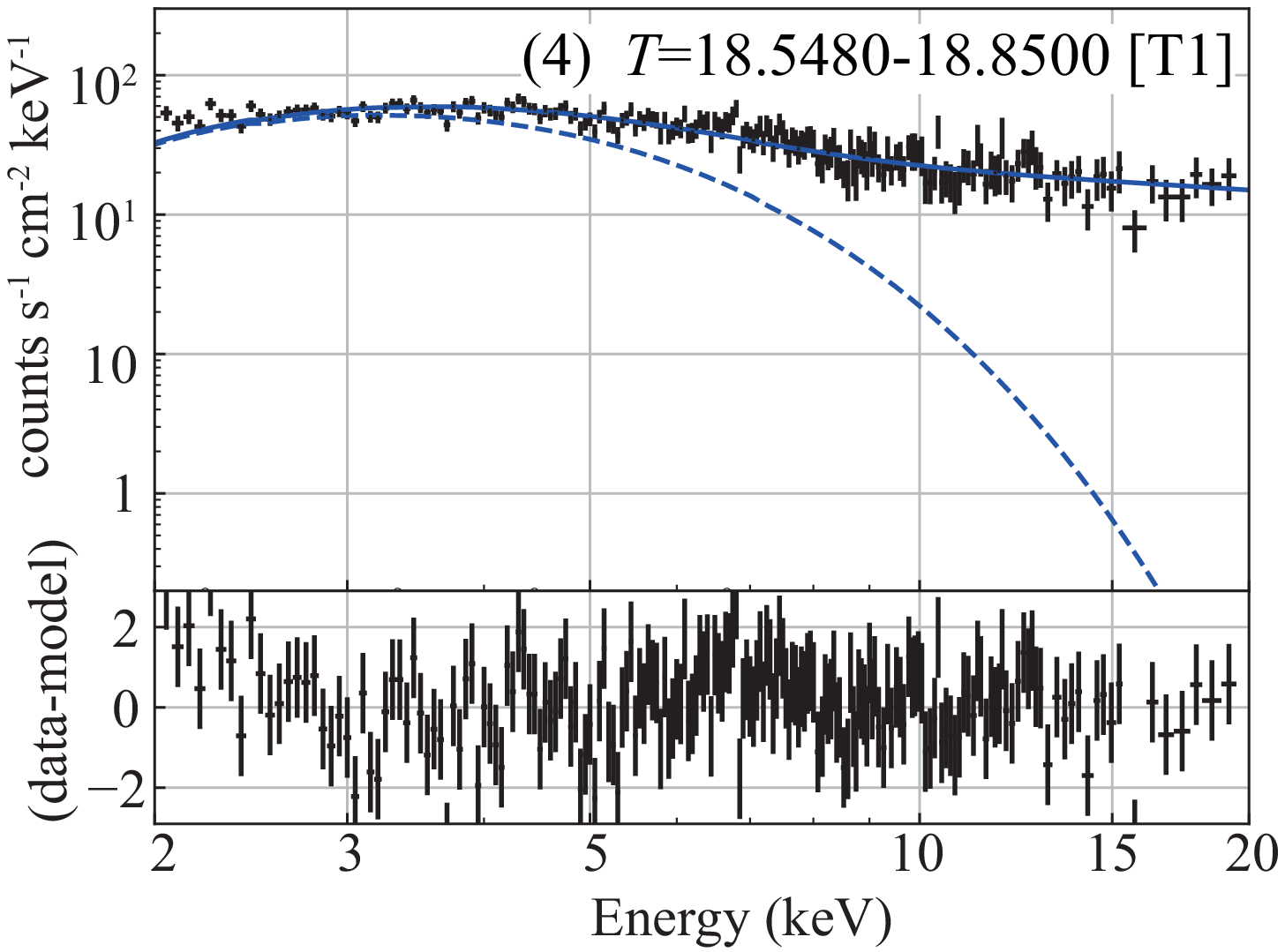}
    \FigureFile(55mm, 55mm){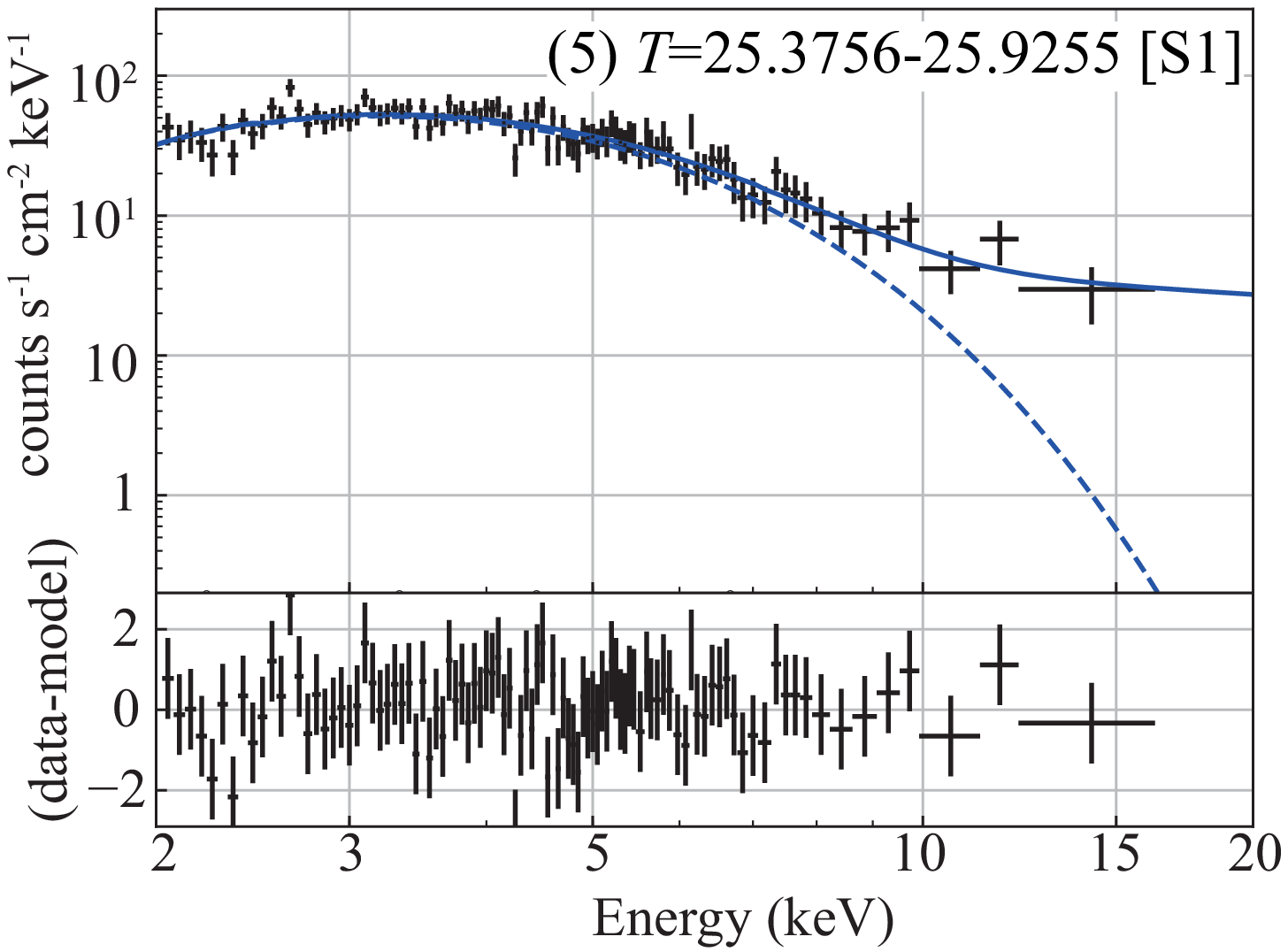}
    \FigureFile(55mm, 55mm){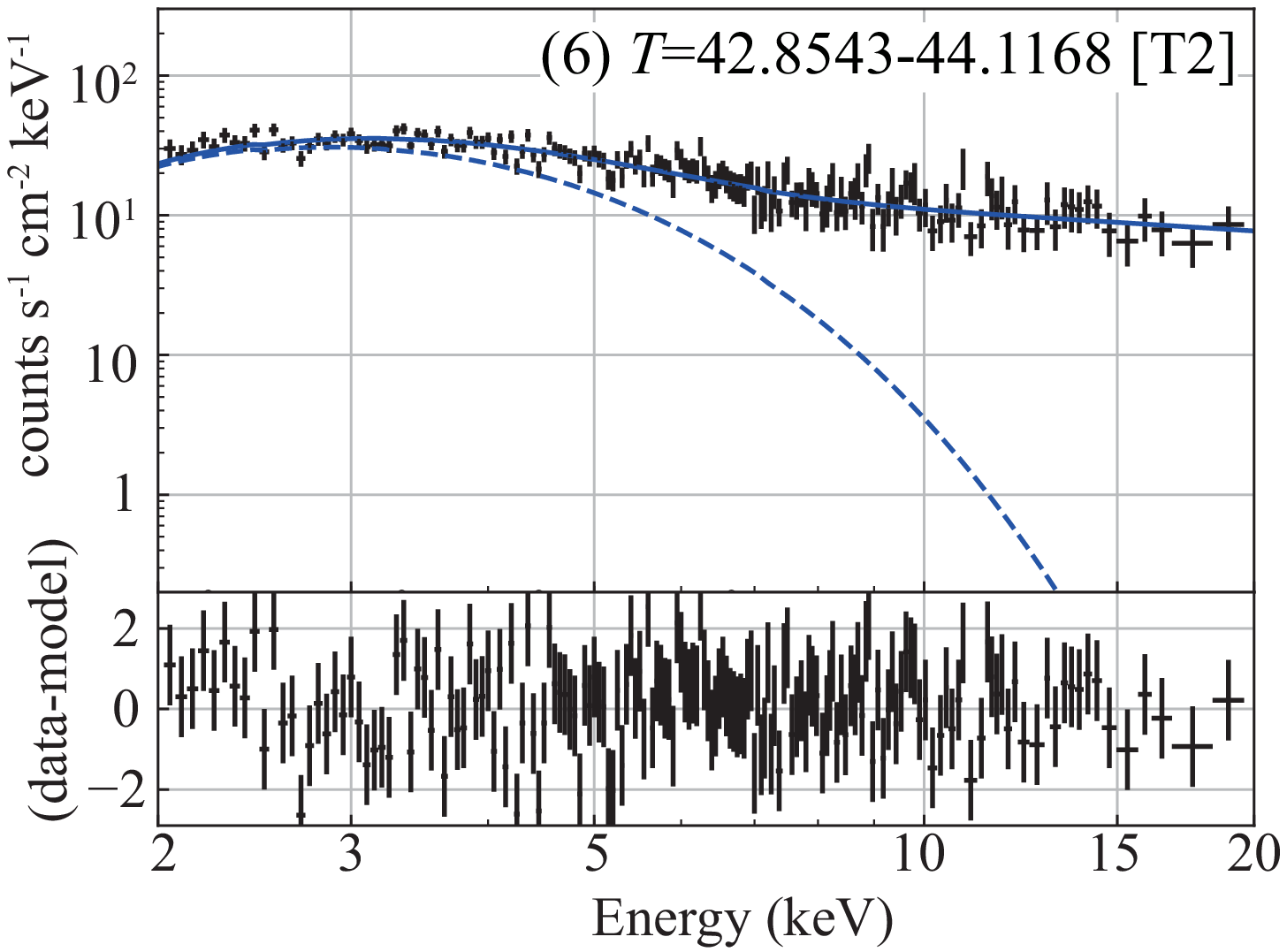}
    \FigureFile(55mm, 55mm){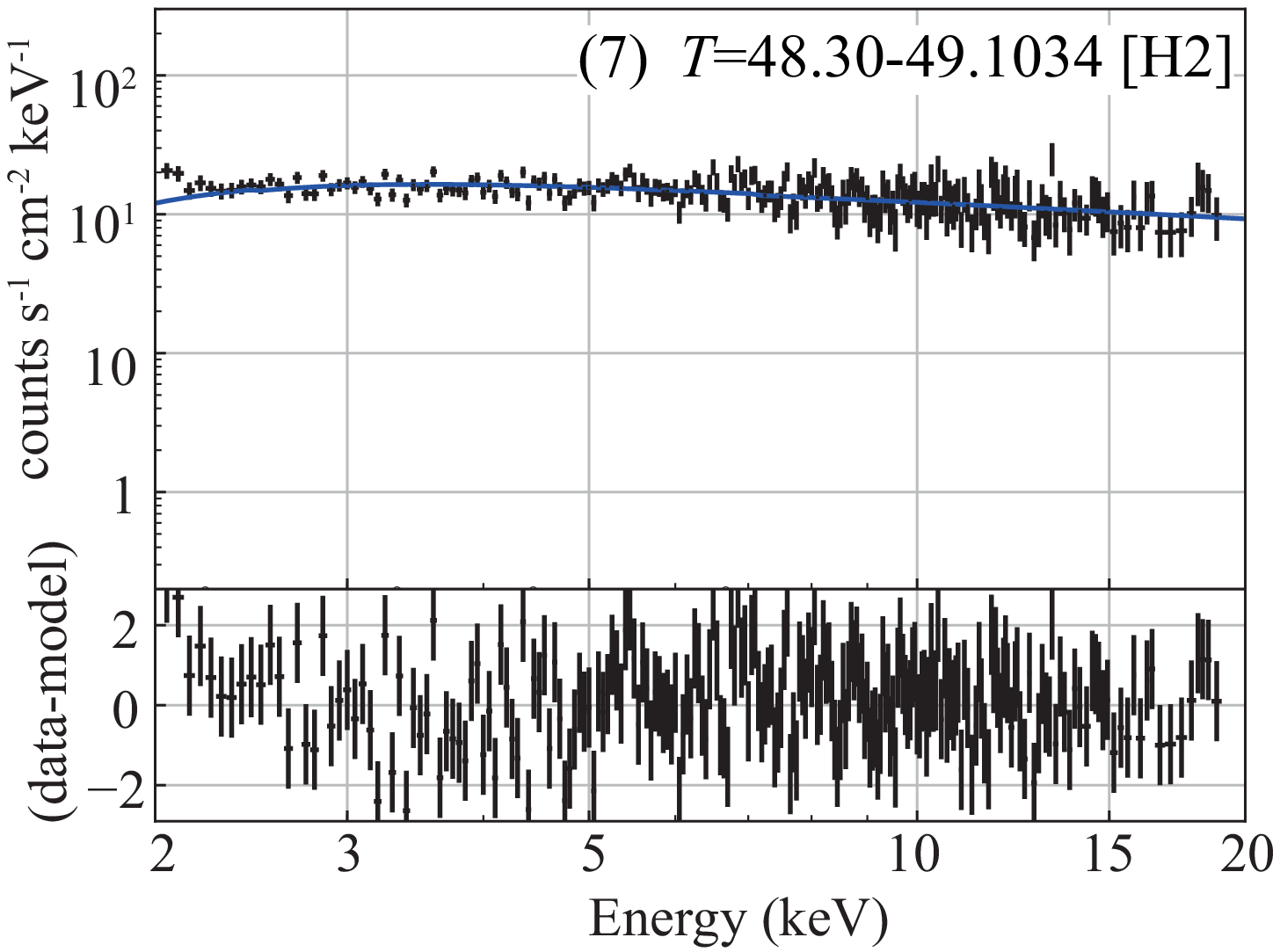}
    \FigureFile(55mm, 55mm){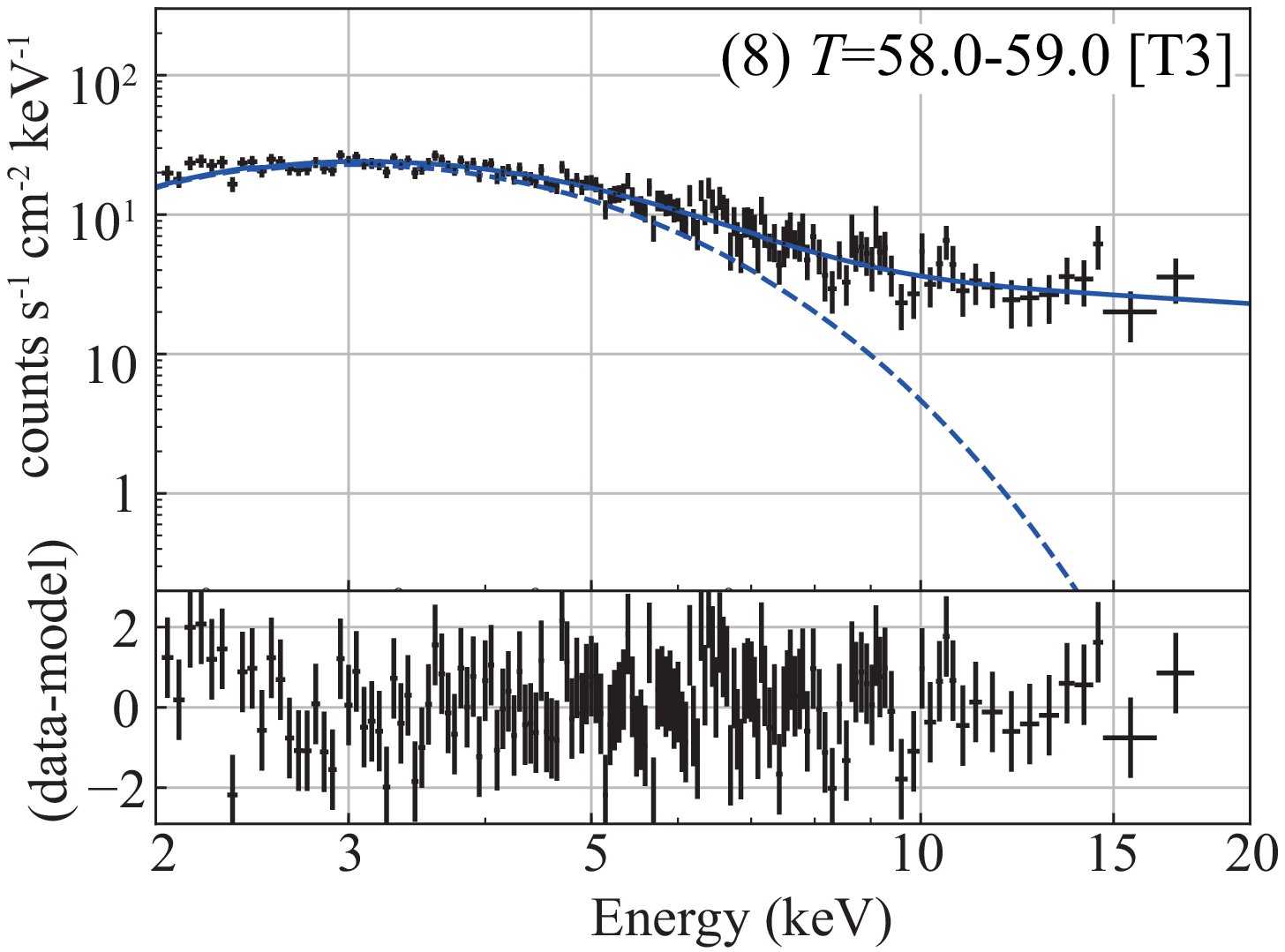}
    \FigureFile(55mm, 55mm){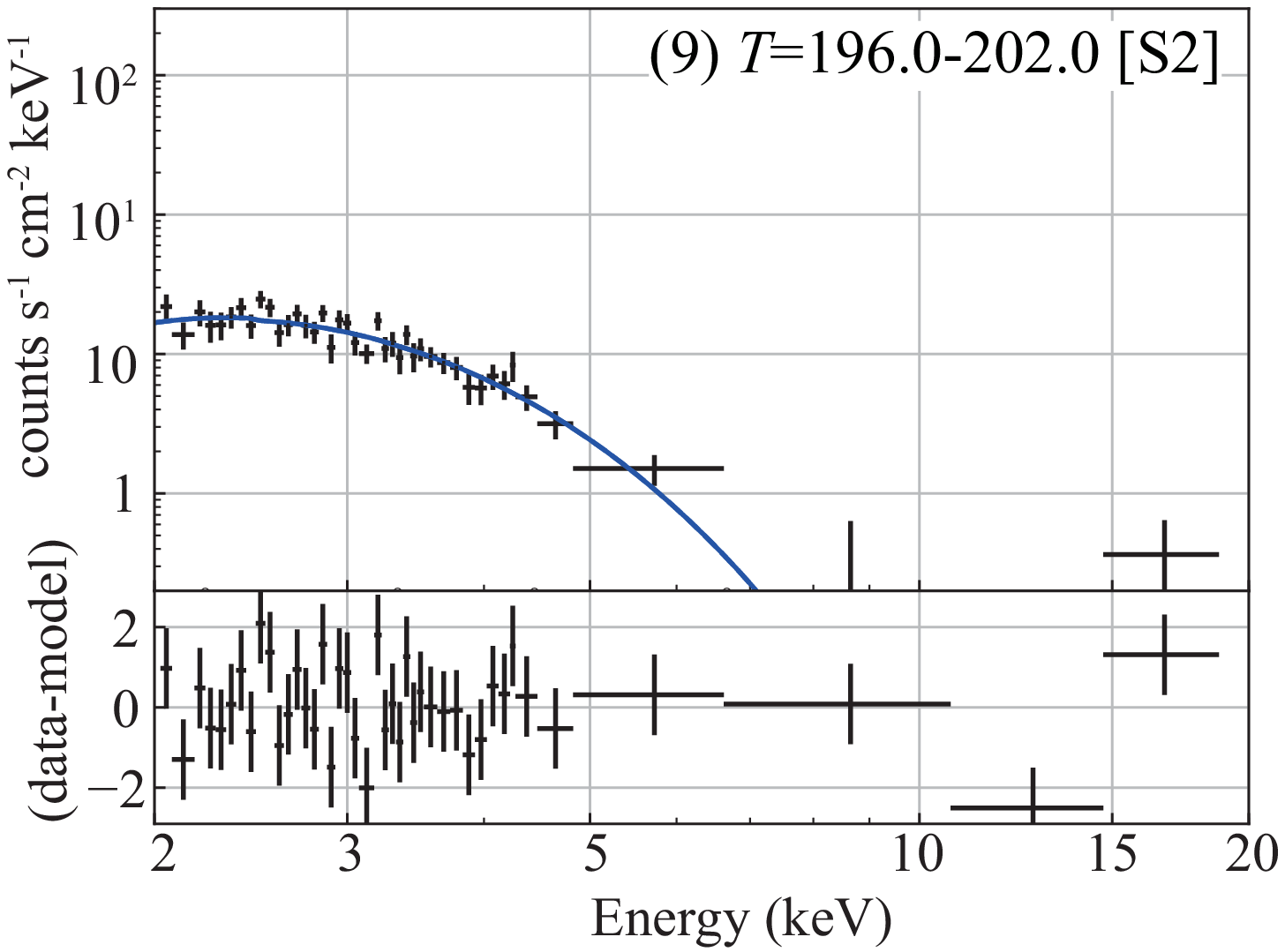}
  \end{center}
  \caption{
Representative $\maxi$/GSC spectra in the individual phases (given in square brackets).
The solid line shows the total model spectrum, and the dashed line indicates the intrinsic disk contribution.
}
\label{fig:specfig}
\end{figure*}

\begin{figure*}
  \begin{center}
    \FigureFile(170mm,170mm){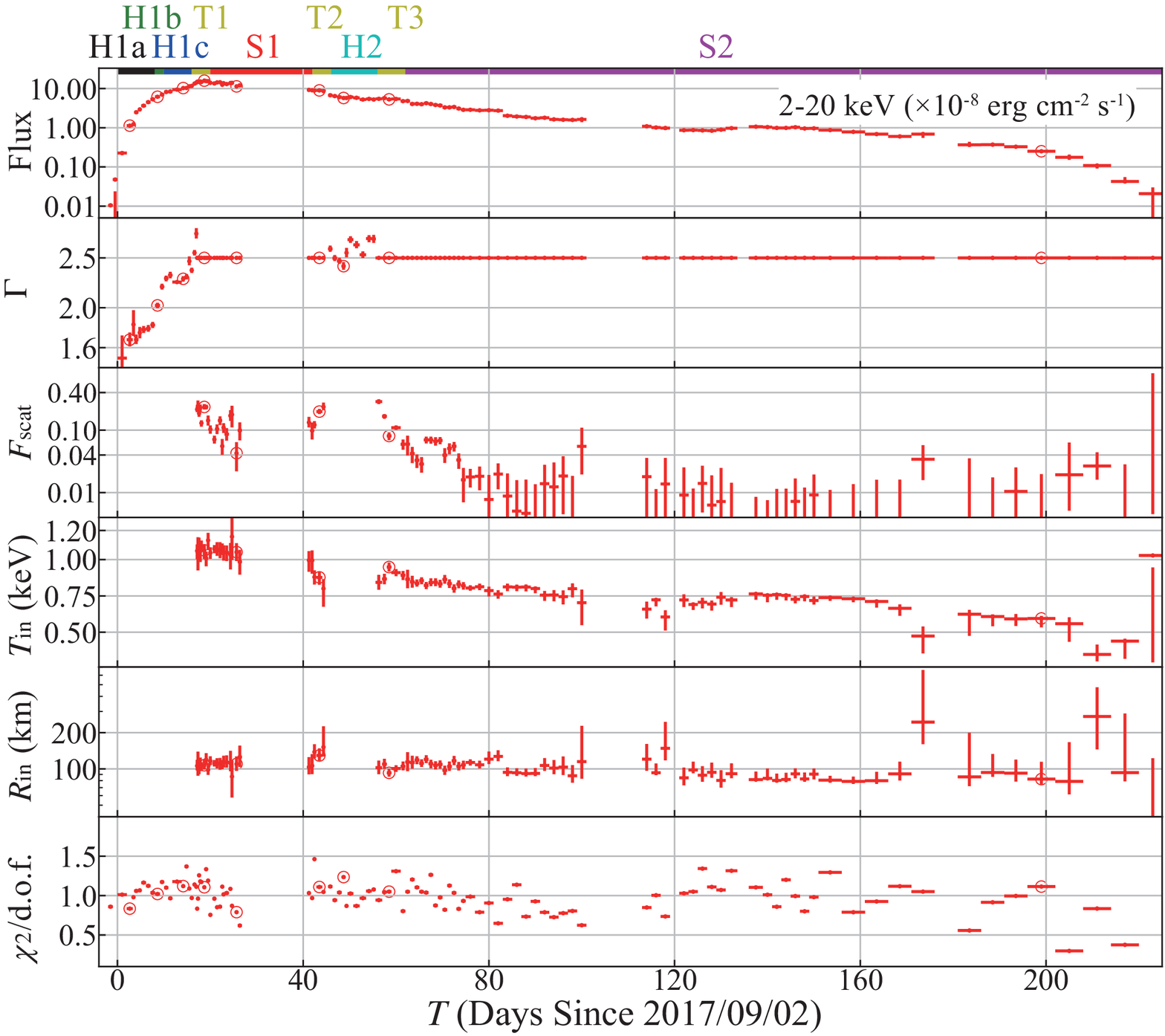}
  \end{center}
  \caption{
  Time evolution of the spectral parameters derived from the model fits 
  to the 2--20 keV $\maxi$/GSC data.
   }
\label{fig:trend_fitpars}
\end{figure*}

Figure~\ref{fig:specfig} and Table~\ref{tab:spec_phase} 
present the spectra and the best-fit models for 
representative spectra in the individual phases.
Analyzing all the spectra in the same say, we have obtained 
successful fits from most of them.
The long-term evolution of the derived best-fit parameters is given 
in figure~\ref{fig:trend_fitpars}. 
For $T$=16--25, 42--45, and 55--202, the $\Rin$ value,
calculated assuming a distance $D$=10 kpc, stayed fairly constant, 
in spite of significant variations of the flux, 
Comptonized fraction, and the inner disk temperature.
In Contrast, $\Rin$ increased and varied significantly 
for $T=$45--53, where rapid spectral hardening occurred.
The results generally confirm our state assignments employed so far;
H1 and H2 to the hard state and S1 and S2 to the soft state.

\if 1

cp -r LC/MAXIJ1535_LC.pdf ~/Documents/MAXIJ1535/trunk/figure/
cp -r LC/MAXIJ1535_HID.pdf ~/Documents/MAXIJ1535/trunk/figure/
cp -r LC/MAXIJ1535LC_trans1.pdf ~/Documents/MAXIJ1535/trunk/figure/
cp -r GSCSPEC3/specpar_stack.pdf ~/Documents/MAXIJ1535/trunk/figure/

cp -r GSCSPEC3/20170904_0000/20170904_0000_spec.pdf ~/Documents/MAXIJ1535/trunk/figure/spec1_H1a_20170904_0000.pdf
cp -r GSCSPEC3/20170910_0147/20170910_0147_spec.pdf ~/Documents/MAXIJ1535/trunk/figure/spec2_H1b_20170910_0147.pdf
cp -r GSCSPEC3/20170915_1938/20170915_1938_spec.pdf ~/Documents/MAXIJ1535/trunk/figure/spec3_H1c_20170915_1938.pdf
cp -r GSCSPEC3/20170920_1309/20170920_1309_spec.pdf ~/Documents/MAXIJ1535/trunk/figure/spec4_T1_20170920_1309.pdf
cp -r GSCSPEC3/20170927_0900/20170927_0900_spec.pdf ~/Documents/MAXIJ1535/trunk/figure/spec5_S1_20170927_0900.pdf
cp -r GSCSPEC3/20171014_2030/20171014_2030_spec.pdf ~/Documents/MAXIJ1535/trunk/figure/spec6_T2_20171014_2030.pdf
cp -r GSCSPEC3/20171020_0712/20171020_0712_spec.pdf ~/Documents/MAXIJ1535/trunk/figure/spec7_H2_20171020_0712.pdf
cp -r GSCSPEC3/20171030_0000/20171030_0000_spec.pdf ~/Documents/MAXIJ1535/trunk/figure/spec8_T3_20171030_0000.pdf
cp -r GSCSPEC3/20180317_0000/20180317_0000_spec.pdf ~/Documents/MAXIJ1535/trunk/figure/spec9_S2_20180317_0000.pdf

python ana_xspec_simpl.py GSCSPEC3 20170904_0000
python ana_xspec_simpl.py GSCSPEC3 20170910_0147
python ana_xspec_simpl.py GSCSPEC3 20170915_1938
python ana_xspec_simpl.py GSCSPEC3 20170920_1309
python ana_xspec_simpl.py GSCSPEC3 20170927_0900
python ana_xspec_simpl.py GSCSPEC3 20171014_2030
python ana_xspec_simpl.py GSCSPEC3 20171020_0712
python ana_xspec_simpl.py GSCSPEC3 20171030_0000
python ana_xspec_simpl.py GSCSPEC3 20180317_0000

((c.c**2*100*u.km)/(6*c.G)/c.M_sun)

\fi

\begin{table*}
  \caption{The best-fit model parameters obtained from the {\maxi}/GSC spectra.}
  \begin{center}         
  \begin{tabular}{ccccccccc}
  \hline\hline
time[phase]   & Flux\footnotemark[$*$]  &   $\Gamma$  &  $\fscat$\footnotemark[$\dagger$]  & $\Tin$ & $\Rin$\footnotemark[$\ddagger$] & $\chi^2$/d.o.f.  \\ 
  \hline\hline
    2.0000--3.2865[H1a]     &       1.33$\pm0.05$       &      1.58$\pm0.06$       &             -             &            -             &               -               &    1.00(182)   \\
    8.0745--9.1666[H1b]    &       6.08$\pm0.09$       &       2.00$\pm0.03$       &             -             &            -             &              -               &    1.02(284)   \\
   13.8187--14.5000[H1c]   &       10.14$\pm0.15$      &       2.26$\pm0.03$       &             -             &            -             &              -               &    1.12(256)   \\
   18.5480--18.8500[T1]  &  15.62$^{+0.31}_{-0.30}$  &  2.50 (fixed)  &   0.21$^{+0.03}_{-0.02}$  &  1.00$^{+0.04}_{-0.06}$  &   99.15$^{+13.49}_{-8.09}$   &    1.106(193)    \\
   25.3756--25.9255[S1]   &  10.87$^{+0.43}_{-0.50}$  &  2.50 (fixed)  &       0.02$\pm0.02$       &      0.99$\pm0.06$       &  96.32$^{+16.83}_{-13.59}$   &    0.790(96)    \\
   42.8543--44.1168[T2]   &       8.76$\pm0.20$       &  2.50 (fixed)  &       0.18$\pm0.02$       &  0.83$^{+0.04}_{-0.05}$  &  116.97$^{+18.99}_{-12.22}$  &    1.110(167)   \\
   48.30--49.1034[H2]   &       5.64$\pm0.10$       &       2.38$\pm0.04$       &             -             &            -             &              -               &    1.236(231)   \\
   58.0--59.0[T3]   &   5.19$^{+0.12}_{-0.11}$  &       2.50 (fixed)       &       0.07$\pm0.01$       &      0.92$\pm0.03$       &    86.14$^{+7.08}_{-7.07}$    &    1.05(144)   \\
  196.0--202.0 [S2]  &   0.24$^{+0.02}_{-0.01}$  &  0.00$^{+-2.50}_{-2.50}$  &  0.00$^{+0.02}_{--0.00}$  &  0.53$^{+0.02}_{-0.06}$  &   74.37$^{+32.09}_{-7.92}$   &    1.115(39)    \\
   \hline
  \end{tabular}
  \label{tab:spec_phase} 
  \end{center}
\footnotemark[$*$]{2--20 keV X-ray flux in the units of $\times$ 10$^{-8}$ $\erg$)} 
\footnotemark[$\dagger$]{Scattering fraction.} 
\footnotemark[$\ddagger$]{Inner radius estimated from the normalization of {\tt diskbb} by assuming the distance and inclination of 10 kpc and 0$^\circ$, respectively.}
\end{table*}

\section{Discussion}\label{sec:discussion}
\subsection{Overall properties of the outburst}\label{sec:discuss-overall}
We have observed a dramatic outburst of MAXI J1535$-$571, with a 
peak X-ray intensity of $\sim$5~Crab, which is the 7th highest 
value among those of BH candidates observed so far. 
The source reached the outburst peak more than two weeks after the discovery, 
which is relatively long compared 
with the typical flux-rise time scales in BH candidates. 
Assuming a typical distance of $D$=10 kpc and an isotropic emission, 
the 2-20 keV peak luminosity is obtained as $\sim$ 2.0 $\times$ 10$^{39}$ erg s$^{-1}$, 
which is comparable to the Eddington luminosity of a 10 $\msolar$ BH where $\msolar$ is the solar mass.
Further discussion continues later on.

The evolution of the 2--20 keV flux is roughly characterized 
by a simple function of time, with a linear increase until $T=18.5$ 
followed by an exponential decay and a re-flaring around $T=130$. 
Meanwhile, its energy spectra exhibited sequential changes as shown in 
figures~\ref{fig:specfig}. In the following, we 
examine the spectral evolution. 

At the beginning, MAXI J1535$-$571 brightened monotonically, accompanied by an 
increase of the photon index from $\Gamma \sim 1.5$ to $\sim 2.0$ (H1a through H1c),
and then rapidly softened through the transition T1, to reach the soft state S1.
Subsequently, it gradually declined  in $\sim$ 180 d with a decrease 
of $\Tin$ from $\sim$1 keV to 0.55 keV.
However, during that course, it suddenly exhibited a spectral hardening (T2), 
stayed in a presumably hard state (H2) for about ten days, and jumped back (T3)
to the soft state (S2) to continue its decline.
In H1c [$T$=10--16; figure~\ref{fig:specfig} (3)] and 
H2 [$T$=46-56; figure~\ref{fig:specfig} (7)], the source stayed for fairly long times, 
in a state of flat spectrum with $\Gamma\simeq$2.0 (equivalent to HR of 0.2--0.4),
even though the HR changed widely from $\sim$0.8 down to 0.01.
The source declined rapidly after $T \sim$200 \citep{2018ATel11568....1N}. 
After T=220, the source became undetectable with the MAXI/GSC, 
whith a typical detection limit of 3.0$\times 10^{-10} \erg$ in 2--20 keV.
Then, a spectral hardening was reported in the Swift/XRT observation at 
$T$=242 \citep{2018ATel11611....1R}. The overall behavior 
in 
the spectral evolution of MAXI J1535$-$571 
supports the assumption that it is a BHB, 
although there are some characteristics unique 
to this source, including as described below, the transition 
luminosity from the soft state back to the hard state.

In figure~\ref{fig:trend_fitpars}, the latest MAXI/GSC data point 
($T$=214--220) gives the lowest 2--20 keV 
X-ray flux during the decay phase, 
$\sim 0.04 \times 10^{-8}$ erg cm$^{-2}$ s$^{-1}$,  
which corresponds to 0.25\% of the peak flux (15.7 
$\times 10^{-8}$ erg cm$^{-2}$ s$^{-1}$). 
On that occasion, the source still remained in the 
soft state. At $T$=242, however, the Swift spectrum 
was described with a power-law model with 
$\Gamma \sim$1.6, meaning that the source finally returned to the hard state.
The 3--7 keV flux at that time, 
2.2$\times$ 10$^{-12}$ $\erg$ cm$^{-2}$ s$^{-1}$, 
is converted to the 2--20 keV flux of $\sim$6.4 
$\times$10$^{-12}$ $\erg$ cm$^{-2}$ s$^{-1}$,  
or 4$\times$10$^{-5}$ of the peak flux.  
This is unusually low compared with typical thresholds, 
$\sim$ 10\% of the peak flux, observed in outburst 
of BHBs, such as XTE~J1752--223 
\citep{2012PASJ...64...13N}, 
MAXI~J1659--152 \citep{2012PASJ...64...32Y},
and MAXI~J1910--057 \citep{2014PASJ...66...84N}.
This means that the transition in MAXI J1535$-$571 
to the hard state occurred at much lower luminosity 
than in other BHBs, or that its peak luminosity was much 
higher than those of the others, both in terms of Eddington ratio.

\subsection{Interpretation of the behavior on the HID}\label{sec:discuss-interpretation}

As in many black hole transients, we can clearly see hysteresis patterns in 
the HID (figure\ref{fig:HID}), on which the source roughly drew a counter-clockwise track.
However, the track is split into two parts by an excursion 
to the hard state H2 as T2$\rightarrow$H2$\rightarrow$T3, which occurred when the luminosity decreased 
by a factor of $\sim 3$, from the peak value in T1 which is considered to be 
close to Eddington limit (section\ref{sec:discuss-constraint}). 
Interestingly, the softest phase in the upper section (S1) has a 
spectrum similar to the softest spectrum in the lower section (S2), 
both dominated by the disk blackbody component, with similar $\Rin$ 
but different $\Tin$. 
However, as is clear by comparing spectra (5) and (9) in figure~\ref{fig:specfig} 
and inspecting figure~\ref{fig:trend_fitpars}, 
the power-law contribution (represented by $\fscat$) is considerably higher in S1, 
than in S2 when $\fscat$ was a few percent like in typical BHBs in the soft state.
In addition, the power-law component in S1 exhibited the noticeable 
variability (figure~\ref{fig:lightcurve}b, section~\ref{sec:lightcurve}).
Therefore, S1 and S2 could be somewhat different.

A pattern on the HID, similar to that of MAXI J1535$-$571, is found in 
figure 4 of \citet{2016ASSL..440...61B} as ``generic HID''. 
According to these authors, our H1c and H2 could be classified as the hard-intermediate state, 
while S1 as the soft-intermediate state. 
Clues to a more conclusive classification of the spectral states 
would be provided by spectral shapes above 10 keV, and the short-term flux 
variability; these are left as future studies using the present data.

In H1c and T1, the 2--8 keV and 15--50 keV fluxes showed a small, 
repetitive fluctuations on a timescale of $\sim$one day, 
with a sign of an anti-correlation (figure~\ref{fig:lightcurve}b; section\ref{sec:lightcurve}). 
The time averaged spectra in H1c have a fairly flat 
profile, characterized by a single power-law component. 
Thus, the anti-correlated flux variation would be 
produced by variations of the spectral index 
with a pivot energy at $\sim$ 10 keV. However, the low statistics of the MAXI 
data hampered us to detect significant spectral differences at different 
phases of the fluctuation.

An intensity variation on a similar timescale was observed previously from 
MAXI J1659$-$152, but it has several 
different properties from that of MAXI J1535$-$571. The variation in 
MAXI J1535$-$571 was observed near the flux peak, whereas that 
of MAXI J1659$-$152 was observed in the decaying phase of an outburst. 
Moreover, the case of MAXI J1659$-$152 is characterized by positive 
correlations between the soft and hard signals, and hence it was 
suggested to be produced by precession 
of the accretion disk induced by a 3:1 resonance of its orbital period \citet{2013A&A...552A..32K}. 
The origin of the peculiar behavior in 
MAXI J1535$-$571 is hence still unclear, but could be associated 
with some changes in the structure of the Comptonized corona, 
in response to fluctuations of the mass accretion rate. 
A slight increase/decrease of the mass accretion rate 
would cause an increase/decrease of the number of seed photons 
for Comptonization in the corona, and thereby the corona could be cooled/heated, 
leading to a change in the Comptonized continuum (e.g., \cite{2014ApJ...789..100S}). 

\subsection{Constraint on the black-hole mass}\label{sec:discuss-constraint}

For $T$=16--25 (T1 to S1) and $T$=55--202 (T3 to S2) in figure~\ref{fig:trend_fitpars}, $\Rin$ remained relatively constant 
at an average of 101 ($D/10$ kpc) $(\cos i/ cos0)^{-1/2}$ km, despite the large flux change over nearly two orders of magnitude.
Therefore, in these periods, the inner edge of the standard disk 
is likely to have reached the innermost stable circular orbit (ISCO).
Now that the ISCO information has been obtained as above,
let us constrain the black hole mass $M_{\rm BH}$ of MAXI~J1535$-$571,
utilizing a $D$ versus $M_{\rm BH}$ diagram in figure 5.
There, a pair of solid black lines were obtained by identifying the observed disk radius, 
$\Rin$ = 101 ($D/10$ kpc) $(\cos i/ cos0)^{-1/2}$ km, 
with the ISCO of a Schwarzschild BH, namely, three times the 
Schwarzschild radius. 
Here, $i$ is the disk inclination, which is assumed to be $i=0^\circ$ and $60^\circ$. 
The red dashed lines represent the condition
that the outburst peak luminosity $\lpeak$
corresponds to 10, 20, 50, and 100\% of $\ledd$. 
In order for $M_{\rm BH}$ to fall in a typical mass range
of $5-15~M_\odot$ \citep{2006csxs.book..157M},
the ratio $L_{\rm peak}/\ledd$ should be 
higher than, say, several tens percent.

In the above estimates, we assumed a Schwarzschild black hole,
which may not be realistic.
If the BH is spinning considerably,
as suggested by \citet{2018ApJ...852L..34X} 
by modeling the reflection feature in the NuSTAR spectrum of this object,
the ISCO radius should decrease, 
down to $\sim 0.5$ Schwarzschild radii at the extreme.
Furthermore, the disk spectrum would be 
more strongly affected by relativistic effects.
However, the NuSTAR data were acquired at $T=5.78$,
or at a phase H1a in our data,
when the object was clearly in the hard state.
It is not clear to us whether the disk in the hard state
can really get so close ($<2.1$ times the ISCO radius; 
\cite{2018ApJ...852L..34X}) to the central BH,
considering that the disk in the hard state is usually 
observed to truncate at radii considerably larger than ISCO \citep{2011PASJ...63S.785S}.
Since the the MAXI/GSC data are limited in statistics,
applying a relativistic disk emission model may not give meaningful constraints on the BH spin.
This is also left for our future studies.

\begin{figure}
  \begin{center}
    \FigureFile(70mm,70mm){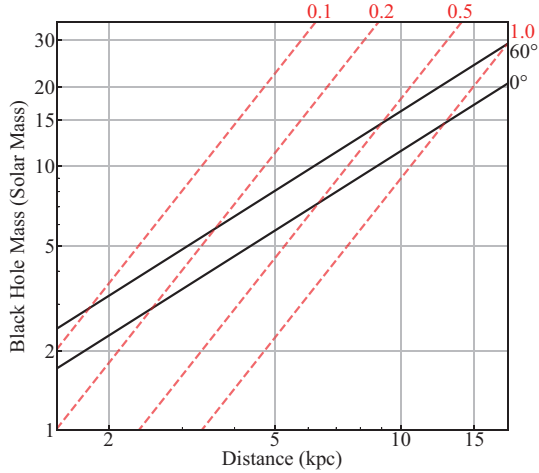}
  \end{center}
  \caption{
 A distance-mass diagram of MAXI~J1535$-$571, with two sets of 
 observational constraints. 
 Black lines show those obtained by identifying the observed values 
 of $\Rin$ with ISCO of Schwarzschild BHs, 
 whereas red dashed lines give constraints by estimating the Eddington 
 limits from the observation.
 The associated numbers in red indicate the assumed $\lpeak$/$\ledd$ ratio.
   }\label{fig:dist_vs_Leddratio}
\end{figure}

\section{Summary and conclusions}\label{sec:summary}

Using the MAXI/GSC, we monitored the X-ray spectral evolution of the 
particularly bright black hole binary MAXI~J1535$-$571,
from the onset of its outburst until the source faded below 
the detection limit at $T>$220.
\begin{enumerate}
\item On the hardness vs 2-20 keV intensity diagram, the source showed
a hysteresis behavior, by exhibiting a few transitions between the hard 
state and the soft state. 
During the decline phase in the soft state at the 30--40\% of the peak
luminosity it made an excursion to relatively hard state for $\sim$10 days. 
\item When the source was in the hard state, prior to the excursion, it exhibited 
quasi-periodic fluctuations on a time scale of $\sim$1-day, with an amplitude of 10--20 \% 
and a clean anti-correlation betreen 2-8 keV and 15 50 keV.
This is likely to be caused by slope changes in the power lat component,
which dominated the spectrum at that epoch.
\item The MAXI/GSC energy spectrum can be fitted by a single powerlaw, or a 
Comptomized disk blackbody model.
From the peak to almost the end, the innermost disk temperature and the 2--20 keV 
X-ray flux was observed to decrease from 1.2 to 0.5 keV, and 
from 15.7$\times 10^{-8}\ \erg$ to 0.04$\times 10^{-8}\ \erg$, respectively. 
Meanwhile, the innermost disk radius stayed relatively constant at 
101 $(D/10~\mathrm{kpc})~(\cos i/\cos 0^\circ)^{-1/2}$ km.
\end{enumerate}

\section{acknowledgements}\label{sec:acknowledgements}

This research has made use of {\maxi} data provided by RIKEN, JAXA and
the {\maxi} team.
Part of this work was financially supported by Grants-in-Aid for
Scientific Research 16K17672 (MS) and 17H06362 (YU, TM, HN, and NK) 
from the Ministry of Education, Culture, Sports, Science and Technology (MEXT) of Japan. MS acknowledges
support by the Special Postdoctoral Researchers Program at RIKEN.

\bibliographystyle{apj}
\bibliography{MAXIJ1535}

\begin{thebibliography}{}
\expandafter\ifx\csname natexlab\endcsname\relax\def\natexlab#1{#1}\fi

\bibitem[{{Belloni} \& {Motta}(2016)}]{2016ASSL..440...61B}
{Belloni}, T.~M., \& {Motta}, S.~E. 2016, in Astrophysics and Space Science
  Library, Vol. 440, Astrophysics of Black Holes: From Fundamental Aspects to
  Latest Developments, ed. C.~{Bambi}, 61

\bibitem[{{Britt} {et~al.}(2017){Britt}, {Bahramian}, \&
  {Strader}}]{2017ATel10816....1B}
{Britt}, C.~T., {Bahramian}, A., \& {Strader}, J. 2017, The Astronomer's
  Telegram, 10816

\bibitem[{{Dincer}(2017)}]{2017ATel10716....1D}
{Dincer}, T. 2017, The Astronomer's Telegram, 10716

\bibitem[{Done {et~al.}(2007)Done, Gierli{\'{n}}ski, \& Kubota}]{Done:2007hu}
Done, C., Gierli{\'{n}}ski, M., \& Kubota, A. 2007, The Astronomy and
  Astrophysics Review, 15, 1

\bibitem[{{Gendreau} {et~al.}(2017){Gendreau}, {Arzoumanian}, {Markwardt},
  {Okajima}, \& {Strohmayer}}]{2017ATel10768....1G}
{Gendreau}, K., {Arzoumanian}, Z., {Markwardt}, C., {Okajima}, T., \&
  {Strohmayer}, T. 2017, The Astronomer's Telegram, 10768

\bibitem[{Hiroi {et~al.}(2013)Hiroi, Ueda, Hayashida, Shidatsu, Sato, Kawamuro,
  Sugizaki, Nakahira, Serino, Kawai, Matsuoka, Mihara, Morii, Nakajima, Negoro,
  Sakamoto, Tomida, Tsuboi, Tsunemi, Ueno, Yamaoka, Yoshida, Asada, Eguchi,
  Hanayama, Higa, Ishikawa, Ishikawa, Isobe, Kohama, Kimura, Morihana,
  Nakagawa, Nakano, Nishimura, Ogawa, Sasaki, Sugimoto, Takagi, Usui, Yamamoto,
  Yamauchi, \& Yoshidome}]{Hiroi:2013gt}
Hiroi, K., Ueda, Y., Hayashida, M., {et~al.} 2013, The Astrophysical Journal
  Supplement Series, 207, 36

\bibitem[{{Homan} \& {Belloni}(2005)}]{2005ApSS.300..107H}
{Homan}, J., \& {Belloni}, T. 2005, \apss, 300, 107

\bibitem[{{Kennea}(2017)}]{2017ATel10731....1K}
{Kennea}, J.~A. 2017, The Astronomer's Telegram, 10731

\bibitem[{{Kennea} {et~al.}(2017){Kennea}, {Evans}, {Beardmore}, {Krimm},
  {Romano}, {Yamaoka}, {Serino}, \& {Negoro}}]{2017ATel10700....1K}
{Kennea}, J.~A., {Evans}, P.~A., {Beardmore}, A.~P., {et~al.} 2017, The
  Astronomer's Telegram, 10700

\bibitem[{{Krimm} {et~al.}(2013){Krimm}, {Holland}, {Corbet}, {Pearlman},
  {Romano}, {Kennea}, {Bloom}, {Barthelmy}, {Baumgartner}, {Cummings},
  {Gehrels}, {Lien}, {Markwardt}, {Palmer}, {Sakamoto}, {Stamatikos}, \&
  {Ukwatta}}]{2013ApJS..209...14K}
{Krimm}, H.~A., {Holland}, S.~T., {Corbet}, R.~H.~D., {et~al.} 2013, \apjs,
  209, 14

\bibitem[{{Kubota} {et~al.}(1998){Kubota}, {Tanaka}, {Makishima}, {Ueda},
  {Dotani}, {Inoue}, \& {Yamaoka}}]{1998PASJ...50..667K}
{Kubota}, A., {Tanaka}, Y., {Makishima}, K., {et~al.} 1998, \pasj, 50, 667

\bibitem[{{Kuulkers} {et~al.}(2013){Kuulkers}, {Kouveliotou}, {Belloni},
  {Cadolle Bel}, {Chenevez}, {D{\'\i}az Trigo}, {Homan}, {Ibarra}, {Kennea},
  {Mu{\~n}oz- Darias}, {Ness}, {Parmar}, {Pollock}, {van den Heuvel}, \& {van
  der Horst}}]{2013A&A...552A..32K}
{Kuulkers}, E., {Kouveliotou}, C., {Belloni}, T., {et~al.} 2013, \aap, 552, A32

\bibitem[{{Markwardt} {et~al.}(2017){Markwardt}, {Burrows}, {Cummings},
  {Kennea}, {Marshall}, {Page}, {Palmer}, \& {Siegel}}]{2017GCN.21788....1M}
{Markwardt}, C.~B., {Burrows}, D.~N., {Cummings}, J.~R., {et~al.} 2017, GRB
  Coordinates Network, Circular Service, No.~21792, \#1 (2017), 21788

\bibitem[{{Matsuoka} {et~al.}(2009){Matsuoka}, {Kawasaki}, {Ueno}, {Tomida},
  {Kohama}, {Suzuki}, {Adachi}, {Ishikawa}, {Mihara}, {Sugizaki}, {Isobe},
  {Nakagawa}, {Tsunemi}, {Miyata}, {Kawai}, {Kataoka}, {Morii}, {Yoshida},
  {Negoro}, {Nakajima}, {Ueda}, {Chujo}, {Yamaoka}, {Yamazaki}, {Nakahira},
  {You}, {Ishiwata}, {Miyoshi}, {Eguchi}, {Hiroi}, {Katayama}, \&
  {Ebisawa}}]{2009PASJ...61..999M}
{Matsuoka}, M., {Kawasaki}, K., {Ueno}, S., {et~al.} 2009, \pasj, 61, 999

\bibitem[{{McClintock} \& {Remillard}(2006)}]{2006csxs.book..157M}
{McClintock}, J.~E., \& {Remillard}, R.~A. 2006, In: Compact Stellar X-Ray
  Sources, ed. W. H. G., Lewin, \& M. van der Klis (Cambridge: Cambridge Univ.
  Press), 157

\bibitem[{{Mereminskiy} \& {Grebenev}(2017)}]{2017ATel10734....1M}
{Mereminskiy}, I.~A., \& {Grebenev}, S.~A. 2017, The Astronomer's Telegram,
  10734

\bibitem[{Mitsuda {et~al.}(1984)Mitsuda, Inoue, \& Koyama}]{Mitsuda:1984ul}
Mitsuda, K., Inoue, H., \& Koyama, K. 1984, Publications of the {\ldots}

\bibitem[{{Nakahira} {et~al.}(2014){Nakahira}, {Negoro}, {Shidatsu}, {Ueda},
  {Mihara}, {Sugizaki}, {Matsuoka}, \& {Onodera}}]{2014PASJ...66...84N}
{Nakahira}, S., {Negoro}, H., {Shidatsu}, M., {et~al.} 2014, Publications of
  the Astronomical Society of Japan, 66, 84

\bibitem[{{Nakahira} {et~al.}(2012){Nakahira}, {Koyama}, {Ueda}, {Yamaoka},
  {Sugizaki}, {Mihara}, {Matsuoka}, {Yoshida}, {Makishima}, {Ebisawa},
  {Kubota}, {Yamada}, {Negoro}, {Hiroi}, {Ishikawa}, {Kawai}, {Kimura},
  {Kitayama}, {Kohama}, {Matsumura}, {Morii}, {Nakajima}, {Serino}, {Shidatsu},
  {Sootome}, {Sugimori}, {Suwa}, {Tomida}, {Tsuboi}, {Tsunemi}, {Ueno}, {Usui},
  {Yamamoto}, {Yamazaki}, {Tashiro}, {Terada}, \& {Seta}}]{2012PASJ...64...13N}
{Nakahira}, S., {Koyama}, S., {Ueda}, Y., {et~al.} 2012, Publications of the
  Astronomical Society of Japan, 64, 13

\bibitem[{Nakahira {et~al.}(2013)Nakahira, Ebisawa, Negoro, Mihara, Sugizaki,
  Serino, Suwa, Asada, \& Tomida}]{Nakahira:2013we}
Nakahira, S., Ebisawa, K., Negoro, H., {et~al.} 2013, Journal of Space Science
  Informatics, 2, 29

\bibitem[{{Nakahira} {et~al.}(2017){Nakahira}, {Negoro}, {Mihara}, {Iwakiri},
  {Sugizaki}, {Shidatsu}, {Matsuoka}, {Ueno}, {Tomida}, {Ishikawa}, {Sugawara},
  {Isobe}, {Shimomukai}, {Kawai}, {Sugita}, {Yoshii}, {Tachibana}, {Harita},
  {Muraki}, {Morita}, {Yoshida}, {Sakamoto}, {Serino}, {Kawakubo}, {Kitaoka},
  {Hashimoto}, {Tsunemi}, {Yoneyama}, {Nakajima}, {Kawase}, {Sakamaki}, {Ueda},
  {Hori}, {Tanimoto}, {Oda}, {Tsuboi}, {Nakamura}, {Sasaki}, {Kawai},
  {Yamauchi}, {Hanyu}, {K}, {Hidaka}, {Kawamuro}, \&
  {Yamaoka}}]{2017ATel10729....1N}
{Nakahira}, S., {Negoro}, H., {Mihara}, T., {et~al.} 2017, The Astronomer's
  Telegram, 10729

\bibitem[{Negoro {et~al.}(2016)Negoro, Kohama, Serino, Saito, Takahashi,
  Miyoshi, Ozawa, Suwa, Asada, Fukushima, Eguchi, Kawai, Kennea, Mihara, Morii,
  Nakahira, Ogawa, Sugawara, Tomida, Ueno, Ishikawa, Isobe, Kawamuro, Kimura,
  Masumitsu, Nakagawa, Nakajima, Sakamoto, Shidatsu, Sugizaki, Sugimoto,
  Suzuki, Takagi, Tanaka, Tsuboi, Tsunemi, Ueda, Yamaoka, Yamauchi, Yoshida, \&
  Matsuoka}]{Negoro:2016jb}
Negoro, H., Kohama, M., Serino, M., {et~al.} 2016, Publications of the
  Astronomical Society of Japan, 68, S1

\bibitem[{{Negoro} {et~al.}(2017{\natexlab{a}}){Negoro}, {Kawase}, {Sugizaki},
  {Ueno}, {Tomida}, {Sugawara}, {Isobe}, {Ishikawa}, {Shimomukai}, {Mihara},
  {Serino}, {Iwakiri}, {Shidatsu}, {Matsuoka}, {Kawai}, {Sugita}, {Yoshii},
  {Tachibana}, {Harita}, {Muraki}, {Morita}, {Yoshida}, {Sakamoto}, {Kawakubo},
  {Kitaoka}, {Hashimoto}, {Tsunemi}, {Yoneyama}, {Nakajima}, {Sakamaki},
  {Ueda}, {Hori}, {Tanimoto}, {Oda}, {Tsuboi}, {Nakamura}, {Sasaki}, {Kawai},
  {Yamauchi}, {Hanyu}, {Hidaka}, {Kawamuro}, \&
  {Yamaoka}}]{2017ATel10708....1N}
{Negoro}, H., {Kawase}, T., {Sugizaki}, M., {et~al.} 2017{\natexlab{a}}, The
  Astronomer's Telegram, 10708

\bibitem[{{Negoro} {et~al.}(2017{\natexlab{b}}){Negoro}, {Ishikawa}, {Ueno},
  {Tomida}, {Sugawara}, {Isobe}, {Shimomukai}, {Mihara}, {Sugizaki}, {Serino},
  {Iwakiri}, {Shidatsu}, {Matsuoka}, {Kawai}, {Sugita}, {Yoshii}, {Tachibana},
  {Harita}, {Muraki}, {Morita}, {Yoshida}, {Sakamoto}, {Kawakubo}, {Kitaoka},
  {Hashimoto}, {Tsunemi}, {Yoneyama}, {Nakajima}, {Kawase}, {Sakamaki}, {Ueda},
  {Hori}, {Tanimoto}, {Oda}, {Tsuboi}, {Nakamura}, {Sasaki}, {Kawai},
  {Yamauchi}, {Hanyu}, {Hidaka}, {Kawamuro}, \&
  {Yamaoka}}]{2017ATel10699....1N}
{Negoro}, H., {Ishikawa}, M., {Ueno}, S., {et~al.} 2017{\natexlab{b}}, The
  Astronomer's Telegram, 10699

\bibitem[{{Negoro} {et~al.}(2018){Negoro}, {Mihara}, {Nakahira}, {Yatabe},
  {Takao}, {Matsuoka}, {Kawai}, {Sugizaki}, {Tachibana}, {Morita}, {Sakamoto},
  {Serino}, {Sugita}, {Kawakubo}, {Hashimoto}, {Yoshida}, {Nakajima},
  {Sakamaki}, {Maruyama}, {Ueno}, {Tomida}, {Ishikawa}, {Sugawara}, {Isobe},
  {Shimomukai}, {Ueda}, {Tanimoto}, {Morita}, {Yamada}, {Tsuboi}, {Iwakiri},
  {Sasaki}, {Kawai}, {Sato}, {Tsunemi}, {Yoneyama}, {Yamauchi}, {Hidaka},
  {Iwahori}, {Kawamuro}, {Yamaoka}, \& {Shidatsu}}]{2018ATel11568....1N}
{Negoro}, H., {Mihara}, T., {Nakahira}, S., {et~al.} 2018, The Astronomer's
  Telegram, 11568, 1

\bibitem[{{Palmer} {et~al.}(2017){Palmer}, {Krimm}, \& {Swift/BAT
  Team}}]{2017ATel10733....1P}
{Palmer}, D.~M., {Krimm}, H.~A., \& {Swift/BAT Team}. 2017, The Astronomer's
  Telegram, 10733

\bibitem[{{Russell} {et~al.}(2017{\natexlab{a}}){Russell}, {Altamirano},
  {Tetarenko}, {Sivakoff}, {Neilsen}, {Miller-Jones}, {van den Eijnden}, \&
  {Jacpot Xrb Collaboration}}]{2017ATel10899....1R}
{Russell}, T.~D., {Altamirano}, D., {Tetarenko}, A.~J., {et~al.}
  2017{\natexlab{a}}, The Astronomer's Telegram, 10899

\bibitem[{{Russell} {et~al.}(2018){Russell}, {Altamirano}, {Miller-Jones},
  {Plotkin}, {Tetarenko}, {Sivakoff}, \& {JACPOT XRB
  Collaboration}}]{2018ATel11611....1R}
{Russell}, T.~D., {Altamirano}, S. R.~D., {Miller-Jones}, J.~C.~A., {et~al.}
  2018, The Astronomer's Telegram, 11611, 1

\bibitem[{{Russell} {et~al.}(2017{\natexlab{b}}){Russell}, {Miller-Jones},
  {Sivakoff}, {Tetarenko}, \& {Jacpot Xrb Collaboration}}]{2017ATel10711....1R}
{Russell}, T.~D., {Miller-Jones}, J.~C.~A., {Sivakoff}, G.~R., {Tetarenko},
  A.~J., \& {Jacpot Xrb Collaboration}. 2017{\natexlab{b}}, The Astronomer's
  Telegram, 10711

\bibitem[{{Scaringi} \& {ASTR211 Students}(2017)}]{2017ATel10702....1S}
{Scaringi}, S., \& {ASTR211 Students}. 2017, The Astronomer's Telegram, 10702

\bibitem[{{Shidatsu} {et~al.}(2011){Shidatsu}, {Ueda}, {Tazaki}, {Yoshikawa},
  {Nagayama}, {Nagata}, {Oi}, {Yamaoka}, {Takahashi}, {Kubota}, {Cottam},
  {Remillard}, \& {Negoro}}]{2011PASJ...63S.785S}
{Shidatsu}, M., {Ueda}, Y., {Tazaki}, F., {et~al.} 2011, Publications of the
  Astronomical Society of Japan, 63, S785

\bibitem[{{Shidatsu} {et~al.}(2014){Shidatsu}, {Ueda}, {Yamada}, {Done},
  {Hori}, {Yamaoka}, {Kubota}, {Nagayama}, \& {Moritani}}]{2014ApJ...789..100S}
{Shidatsu}, M., {Ueda}, Y., {Yamada}, S., {et~al.} 2014, \apj, 789, 100

\bibitem[{{Shidatsu} {et~al.}(2017){Shidatsu}, {Nakahira}, {Negoro}, {Ueno},
  {Tomida}, {Ishikawa}, {Sugawara}, {Isobe}, {Shimomukai}, {Mihara},
  {Sugizaki}, {Iwakiri}, {Matsuoka}, {Kawai}, {Sugita}, {Yoshii}, {Tachibana},
  {Harita}, {Muraki}, {Morita}, {Yoshida}, {Sakamoto}, {Serino}, {Kawakubo},
  {Kitaoka}, {Hashimoto}, {Tsunemi}, {Yoneyama}, {Nakajima}, {Kawase},
  {Sakamaki}, {Ueda}, {Hori}, {Tanimoto}, {Oda}, {Tsuboi}, {Nakamura},
  {Sasaki}, {Kawai}, {Yamauchi}, {Hanyu}, {Hidaka}, {Kawamuro}, \&
  {Yamaoka}}]{2017ATel10761....1S}
{Shidatsu}, M., {Nakahira}, S., {Negoro}, H., {et~al.} 2017, The Astronomer's
  Telegram, 10761

\bibitem[{Shimura(1995)}]{Shimura:1995te}
Shimura, T. 1995, The Astrophysical Journal

\bibitem[{Steiner {et~al.}(2009)Steiner, Narayan, McClintock, \&
  Ebisawa}]{Steiner:2009gh}
Steiner, J.~F., Narayan, R., McClintock, J.~E., \& Ebisawa, K. 2009,
  Publications of the Astronomical Society of the Pacific, 121, 1279

\bibitem[{{Tetarenko} {et~al.}(2017){Tetarenko}, {Russell}, {Miller-Jones},
  {Sivakoff}, \& {Jacpot Xrb Collaboration}}]{2017ATel10745....1T}
{Tetarenko}, A.~J., {Russell}, T.~D., {Miller-Jones}, J.~C.~A., {Sivakoff},
  G.~R., \& {Jacpot Xrb Collaboration}. 2017, The Astronomer's Telegram, 10745

\bibitem[{Wilms {et~al.}(2000)Wilms, Allen, \& McCray}]{Wilms:2000en}
Wilms, J., Allen, A., \& McCray, R. 2000, The Astrophysical Journal, 542, 914

\bibitem[{{Xu} {et~al.}(2018){Xu}, {Harrison}, {Garc{\'\i}a}, {Fabian},
  {F{\"u}rst}, {Gandhi}, {Grefenstette}, {Madsen}, {Miller}, {Parker},
  {Tomsick}, \& {Walton}}]{2018ApJ...852L..34X}
{Xu}, Y., {Harrison}, F.~A., {Garc{\'\i}a}, J.~A., {et~al.} 2018, \apj, 852,
  L34

\bibitem[{{Yamaoka} {et~al.}(2012){Yamaoka}, {Allured}, {Kaaret}, {Kennea},
  {Kawaguchi}, {Gandhi}, {Shaposhnikov}, {Ueda}, {Nakahira}, {Kotani},
  {Negoro}, {Takahashi}, {Yoshida}, {Kawai}, \& {Sugita}}]{2012PASJ...64...32Y}
{Yamaoka}, K., {Allured}, R., {Kaaret}, P., {et~al.} 2012, Publications of the
  Astronomical Society of Japan, 64, 32

\end{thebibliography}

\end{document}